\documentclass[prl,10pt,twocolumn,superscriptaddress,aps,floatfix]{revtex4-1}
\usepackage{graphicx}
\usepackage{natbib}
\usepackage{url}
\usepackage[normalem]{ulem}
\usepackage{color}
\usepackage{mathtools,amssymb}
\begin{document}

\title{Evidence of weak superconductivity at the room-temperature grown LaAlO$_{3}$/SrTiO$_3$ interface}
\author{G. E. D. K. Prawiroatmodjo}
\email{guen@nbi.dk}
\affiliation{%
Center for Quantum Devices, Niels Bohr Institute, University of Copenhagen, Universitetsparken 5, 2100 Copenhagen, Denmark.
}%
\author{F. Trier}%
\affiliation{%
Department of Energy Conversion and Storage, Technical University of Denmark, Ris{\o} Campus, 4000 Roskilde, Denmark.
}%
\author{D. V. Christensen}%
\affiliation{%
Department of Energy Conversion and Storage, Technical University of Denmark, Ris{\o} Campus, 4000 Roskilde, Denmark.
}%
\author{Y. Chen}%
\affiliation{%
Department of Energy Conversion and Storage, Technical University of Denmark, Ris{\o} Campus, 4000 Roskilde, Denmark.
}%
\author{N. Pryds}%
\affiliation{%
Department of Energy Conversion and Storage, Technical University of Denmark, Ris{\o} Campus, 4000 Roskilde, Denmark.
}%
\author{T. S. Jespersen}%
\affiliation{%
 Center for Quantum Devices, Niels Bohr Institute, University of Copenhagen, Universitetsparken 5, 2100 Copenhagen, Denmark.
}%
\date{\today}

\begin{abstract}
The two-dimensional electron gas at the crystalline LaAlO$_{3}$/SrTiO$_{3}$ (c-LAO/STO) interface has sparked large interest due to its exotic properties including an intriguing gate-tunable superconducting phase. While there is growing evidence of pronounced spatial inhomogeneity in the conductivity at STO-based interfaces, the consequences for superconductivity remain largely unknown. We study interfaces based on amorphous LAO top layers grown at room temperature (a-LAO/STO) and demonstrate a superconducting phase similar to c-LAO/STO, however, with a gate-tunable critical temperature of $460 \, \mathrm{mK}$, higher than any previously reported values for c-LAO/STO. The dependence of the superconducting critical current on temperature, magnetic field and backgate-controlled doping is found to be consistently described by a model of a random array of Josephson-coupled superconducting domains.
\end{abstract}

\maketitle


Understanding the physics of normal/superconductor hybrid systems have been a subject of active research since the original work of Josephson \cite{Josephson:1962}. Recently, however, driven by theoretical insights \cite{Kitaev:2001, recher:2001} and experimentally enabled by the development of new materials, nanoscale hybrid devices have led to a number of key breakthroughs in quantum transport \cite{Doh:2005, Hofstetter:2009, Mourik:2012}. Strontium titanate (STO) is a wide-gap insulating perovskite oxide with a strong interdependence of structural, magnetic and electronic properties \cite{Fleury:1968, Schooley:1965}. Interfacing STO with other complex oxides, such as lanthanum aluminate (LAO), leads to a two dimensional electron gas with remarkable properties such as high mobility \cite{ohtomo:2004} and gate-tunable superconductivity \cite{reyren:2007, caviglia:2008} coexisting with magnetism \cite{Bert:2012, Li:2011} and strong spin-orbit coupling \cite{Caviglia:2010}. This system therefore provides the right conditions for creating exotic quantum states in a new generation of hybrid devices with electrostatic control \cite{Fidkowski:2013}. In order to exploit this potential, however, a detailed understanding of the nature of the superconducting phase and how it is affected by nearby electrostatic gates is required, and methods are needed for fabricating advanced device geometries.

Recently, the importance of micron scale inhomogeneity for the properties of the two-dimensional electron system in STO-based heterostructures have become evident from direct spatial mapping of the current distribution, the superfluid density, and the electrostatic landscape \cite{Bert:2012, Kalisky:2013, Honig:2013}. Furthermore, signs of phase-coherent superconductivity in the metallic and insulating states were found \cite{richter_interface_2013, Bucheli:2015, Fillis:2016} and attributed to tetragonal domain boundaries in STO. The superconducting transition is commonly described as a two-dimensional system using the Berezinskii-Kosterlitz-Thouless (BKT) model \cite{BKT:1970, BKT:1973, reyren:2007, caviglia:2008}, valid for homogeneous or granular superconducting thin films \cite{Beasley:1979}, however, an alternative model based on percolation of superconducting islands embedded in a metallic background is also found to provide a consistent description \cite{caprara:2013, caprara_effective_2011, biscaras_multiple_2013}. So far, the possible consequences of inhomogeneity on the critical current and its dependence on magnetic field, temperature and electrostatic doping have not yet been considered.

\begin{figure}
\includegraphics[width=8.5cm]{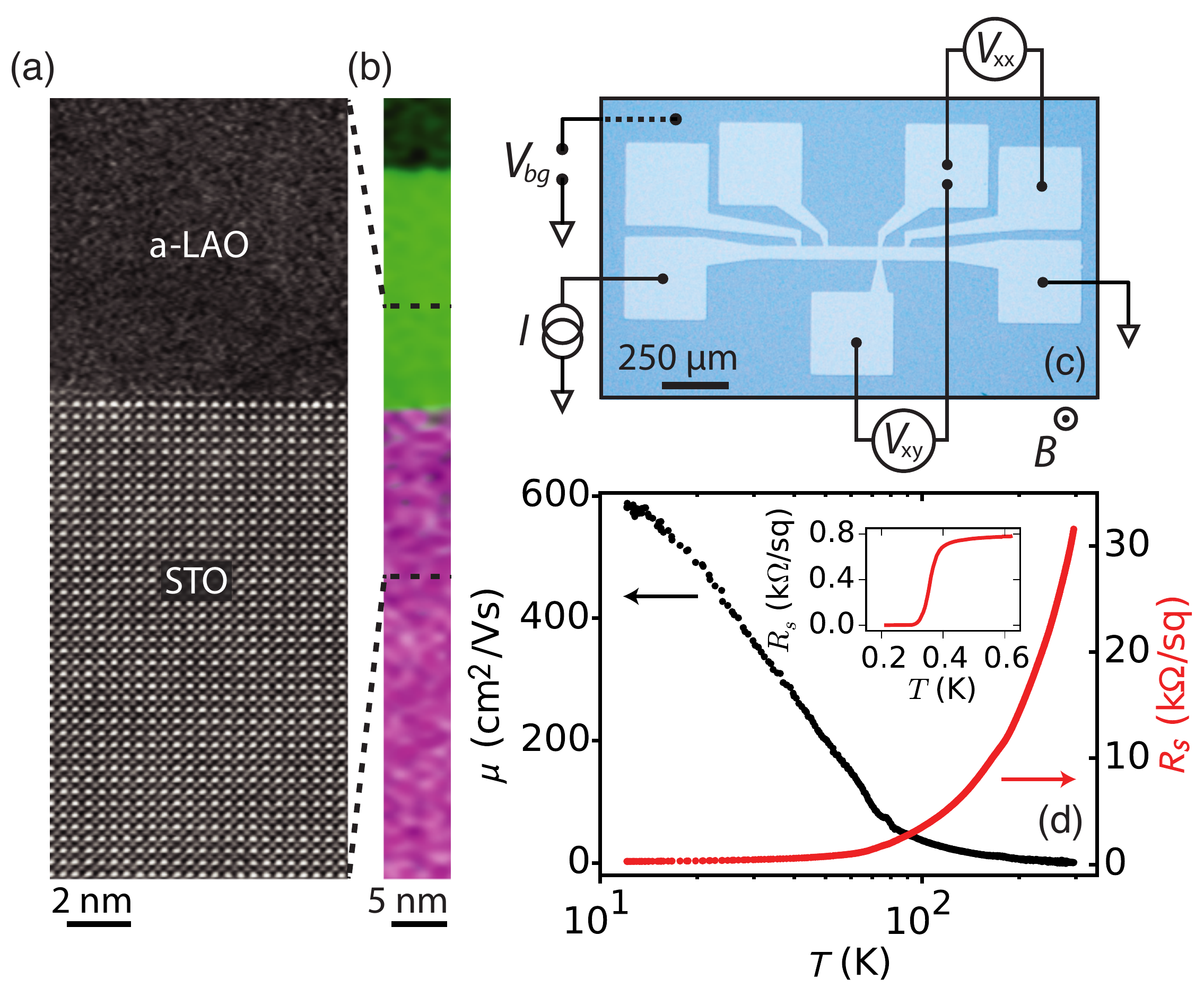}
\caption{(a) HAADF and (b) EELS (La is green and Ti is purple) STEM images showing the amorphous-LAO/STO interface. (c) Optical micrograph of the Hall bar device. The magnetic field $B$ is applied perpendicular to the chip plane and the gate voltage $V_{bg}$ is varied to tune the electrostatic doping of the device. (d) Mobility $\mu$ and sheet resistance $R_{s}$ as a function of the temperature $T$. The superconducting transition at $T = 360 \, \mathrm{mK}$ at $V_{bg} = 0 \, \mathrm{V}$ is shown in the inset.}
\label{fig:hallbar}
\end{figure}

The aim of the present paper is twofold. First we introduce interfaces of STO and room-temperature grown amorphous LAO (a-LAO)\cite{Liu:2013, Chen:2011, Fuchs:2014} to the family of STO-based heterostructures that exhibit exhibit superconductivity. While the doping mechanism leading to conductivity in a-LAO/STO is dominated by oxygen vacancies\cite{Liu:2013, Chen:2011, Fuchs:2014} different from polar discontinuities in c-LAO/STO, the characteristics of superconductivity are found to be similar\cite{reyren:2007, caviglia:2008}.  However, we find that a-LAO/STO exhibits a significantly higher $T_{c}$ than reported for c-LAO/STO and has the added benefit that room temperature growth is compatible with standard semiconductor fabrication processes\cite{Fuchs:2014}. We include a full phase diagram of the critical current dependence on temperature and magnetic field. Secondly, we study and compare the gate-dependence of the superconducting critical current $I_c(V_{bg})$ and critical temperature $T_c(V_{bg})$. Both exhibit a dome-like dependence on $V_{bg}$ however, with a clear shift which provide qualitative evidence a superconducting phase best described to as a random array of superconducting domains \cite{caprara:2013} interconnected by metallic weak links or Josephson junctions (JJs) \cite{mehta:2012}. Such  Josephson junction arrays (JJA) have been shown to undergo a BKT quantum phase transition \cite{BKT:1970, BKT:1973, Vanderzant:1987}, consistent with previous work on c-LAO/STO \cite{reyren:2007, caviglia:2008}. The presence of intrinsic weakly coupled superconducting domains may be a crucial element in the design and study of gate-defined devices at STO-based interfaces \cite{Gallagher:2014}.


Our samples were grown by room-temperature pulsed laser deposition (PLD) and patterned in a Hall bar geometry (width $W= 50 \, \mathrm{\mu m}$, length $L =100 \, \mathrm{\mu m}$) using a LaSrMnO$_3$ hard mask following Ref. \cite{felix_patterning_2014}. The 16 nm LAO top layer is amorphous \cite{Chen:2015} as confirmed by the absence of long-range order in the cross-sectional high-angle annular dark-field (HAADF) scanning transmission electron microscopy (STEM) image and corresponding electron energy loss spectroscopy (EELS) scan in Fig. \ref{fig:hallbar}(a,b) of an unpatterned reference sample. Figure \ref{fig:hallbar}(c) shows an optical image of the final device. The chip was glued to a ceramic chip-carrier using conducting silver paste and the back plane of the chip served as a global electrostatic backgate, tuning the interface carrier density when biased at a voltage $V_{bg}$.

Initial characterization of the device was done by sourcing a current $I$ and measuring the longitudinal and transverse voltages, $V_{xx}$ and $V_{xy}$ while applying an out-of-plane magnetic field $B$ (Fig.\ \ref{fig:hallbar}(c)). The temperature dependence of the sheet resistance $R_s = V_{xx} W /L I$ is shown in Fig.\ \ref{fig:hallbar}(d) confirming the metallic behavior of the sample and the carrier density $n_s = 1/eR_H$ is found from the measured Hall coefficient $R_H= | \partial R_{xy} / \partial B |_{B = 0 \, \mathrm{T}}$. Upon cooling the sample from room temperature, $n_s$ is constant at $0.4 \times 10^{14} \, \mathrm{cm}^{-2}$ until $T = 90 \, \mathrm{K}$ from where it linearly decreases to a value of $0.2 \times 10^{14} \, \mathrm{cm}^{-2}$ at $10 \, \mathrm{K}$. The carrier freeze-out below $\sim 100  \, \mathrm{K}$ is consistent with previous reports on both amorphous and crystalline LAO/STO samples \cite{Huijben:2006, Chen:2011, yunzhong_amorphous_2014}. The mobility $\mu_H = 1/n_s e R_s$ (Fig.\ \ref{fig:hallbar}(d)) increases upon cooling and reaches a value of $\sim 600 \, \mathrm{cm}^{2}/\mathrm{Vs}$ at low temperature. Subsequently, the sample was measured in a dilution refrigerator with a base temperature of $22 \, \mathrm{mK}$. A transition to the superconducting state is observed at $T_c= 360 \, \mathrm{mK}$ (inset to Fig.\ \ref{fig:hallbar}(d)). Here the critical temperature $T_c$ was defined as the temperature where $R_s$ is 50\% of the normal-state resistance $R_N$ at $T = 600 \, \mathrm{mK}$. The transition temperature is comparable to the values ranging from 200 to $300  \, \mathrm{mK}$ reported for c-LAO/STO samples \cite{reyren:2007, caviglia:2008, bert_direct_2011, richter_interface_2013}. 

%
\begin{figure}
\includegraphics[width=8.5cm]{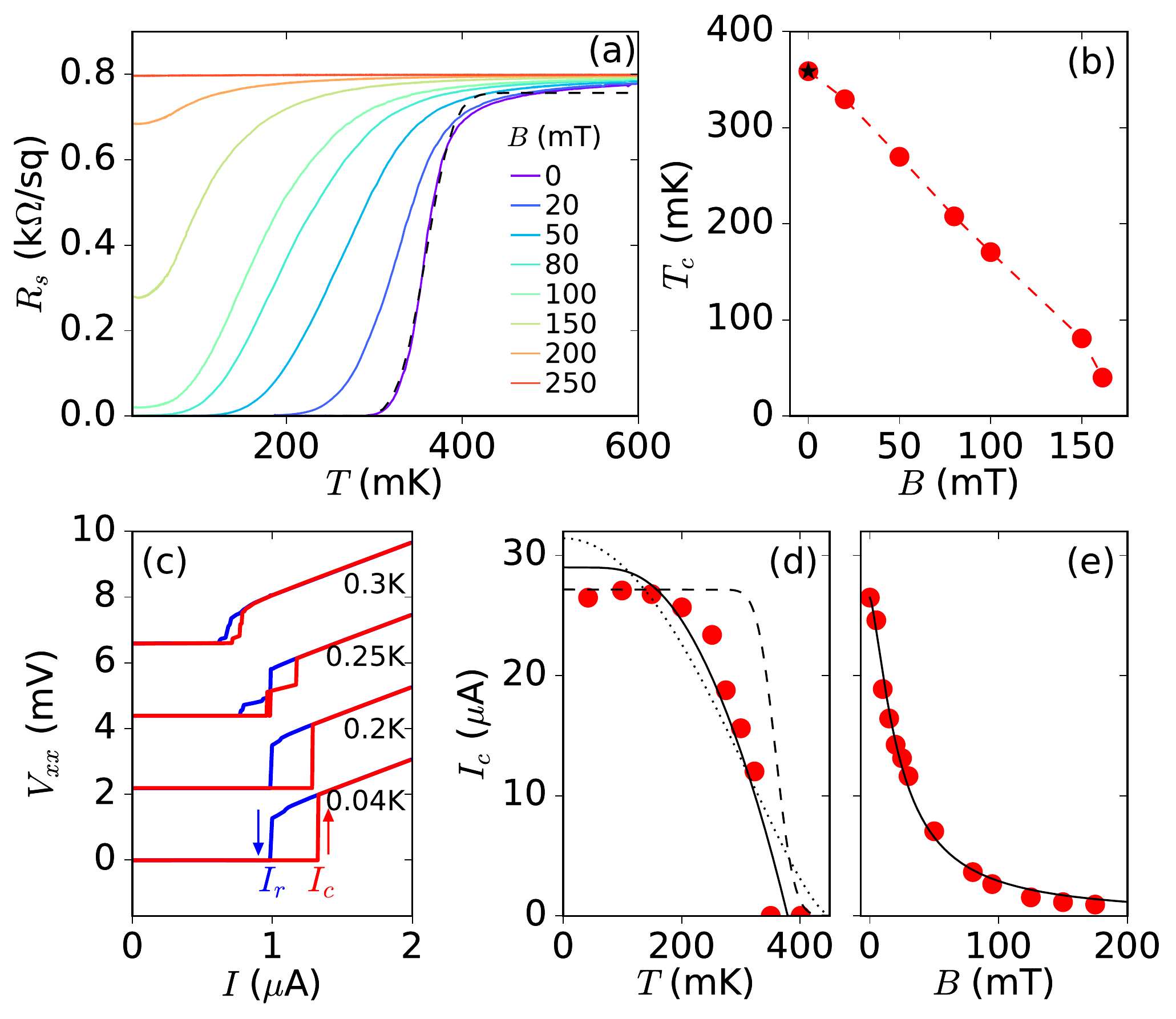}
\caption{\label{fig:ivcurves} (a) $T$-dependence of the $R_S$ at different $B$. The dashed line represents a fit to EMT, the fitted $T_c$ is shown in (b) as a black star. (b) $B$-dependence of $T_{c}$, extracted from (a) with a 50\% criterion, except the lowest $T_c$ which is found by varying the $B$ at $T = 40 \, \mathrm{mK}$. (c) DC measurement of $V_{xx}$ vs. $I$ for various T.  Arrows indicate the sweep direction. Curves are offset by $2.2 \, \mathrm{mV}$. (d) Zero-field T-dependence and $B$-dependence at $30  \, \mathrm{mK}$ (e) of $J_c$. The black lines are fits to theory, as explained in the main text.}
\end{figure}

To study the properties of the superconducting phase, $T_{c}$ was measured at different $B$ sweeping the temperature at a slow $2 \mathrm{mK/min}$ temperature ramp rate to ensure a stable equilibrium situation. Figure \ref{fig:ivcurves}(a) shows $R_s(T)$ for different magnetic field and the resulting $T_c(B)$ is shown in Fig.\ 2(b). Also included in Fig.\ 2(b) is $T_c$ extracted from fitting the $B=0 \, \mathrm{T}$ data to the Effective Medium Theory (EMT) of Ref.\ \cite{caprara:2013} which considers a sample composed of percolating superconducting regions with a Gaussian distribution of transition temperatures with width $\gamma$ and average $\bar{T_c}$. The fit shown by the dashed line in Fig. 2(a) is in good agreement with the experimental curve\footnote{The model does not capture the difference in sharpness of the transition close to $R_N$ and $0 \Omega$ which could be related to a non-symmetric $T_c$ distribution.}. The 50\% transition point (reached at $T=\bar{T_c}$) agrees with the $T_c$ found by fitting EMT theory (black symbol in Fig.\ 2(b)), and extrapolating to $T_c = 0 \, \mathrm{K}$ gives a measure of the upper critical field $B_{c2}(0 \, \mathrm{K}) \approx 180 \, \mathrm{mK}$. This corresponds to a coherence length of $\xi \approx 40 \, \mathrm{nm}$ close to the values found for c-LAO/STO \cite{reyren:2007, ben_shalom_tuning_2010}. Note that the $T_c (B)$ dependence is expected to go to zero $T$ with a vertical tangent \cite{tinkham}, which is not observed due to limits of the measurement. Also, the extracted coherence length depends on the definition of $T_c$ and taking the 5\% or 95\% transition point for $T_c$ results in $\xi(0)$ of $49 \, \mathrm{nm}$ and $35 \, \mathrm{nm}$, respectively.

Four-terminal finite bias $IV$-characteristics are presented in Fig. \ref{fig:ivcurves}(c-e). When increasing the bias current (Fig. 2(c), red trace) the device is initially in the superconducting state and $V_{xx} = 0 \, \mathrm{V}$, but switches abruptly to a resistive state at the critical current $I_{c}$. When reversing the sweep direction (blue trace), the sample returns to the superconducting state at the retrapping current $I_{r} < I_c$. Such hysteretic behavior can be an effect of Joule heating \cite{tinkham_heating_2003, courtois_heating_2008}, however, since the hysteretic behavior is largely unchanged up to $200 \, \mathrm{mK}$, this seems unlikely. Alternatively, hysteretic $IV$-curves are characteristic for an array of Josephson junctions \cite{tinkham, yu:1992} in the underdamped regime. For an individual junction of capacitance $C$ and normal state resistance $R_{JJ}$ the quality factor is $Q = \omega_p R_{JJ} C$, where $\omega_p$ is the plasma frequency, and the underdamped regime $Q \gg 1$ could naturally appear due to the high dielectric constant of STO providing large junction capacitances. In this scenario the multiple switching events to finite resistive states observed at $T >$ $250 \, \mathrm{mK}$ in Fig.\ 2(c) is consistent with an array containing junctions with varying critical currents.

The temperature dependence of the critical current density $J_c = I_c/W$ is shown in Fig. \ref{fig:ivcurves}(d). For $T \lesssim 150 \, \mathrm{mK}$, $J_c$ is constant at $\sim 28 \, \mathrm{mA/m}$ and drops steeply to zero around $350 \, \mathrm{mK}$. The dotted line represents a fit to the Ginzburg-Landau mean field result $J_c^{GL} \propto H_C(0)/\lambda(0) (1-(T/T_c)^2)^{3/2}(1+(T/T_c)^2)^{1/2}$ which describes the critical current in homogeneous superconducting thin films \cite{Skocpol:1974}. Here $H_c(0)$ and $\lambda(0)$ refer to the low temperature value of the critical field and the penetration depth. This model does not describe the data adequately and moreover, taking values $H_c(0) \sim 1000 \, \mathrm{Oe}$ and $\lambda(0) \sim 10 \, \mu\mathrm{m}$ appropriate for bulk STO \cite{Schooley:1965} the model estimates a low field critical current density of $\sim 10 \times 10^{10} \, \mathrm{A/m}^2$. Estimating a superconducting layer thickness of $\sim 10 \, \mathrm{nm}$ \cite{Chen:2011} for our sample this amounts to a density of $10 \, \mathrm{A/m}$ ie., three orders of magnitude larger than what we measure. The $T$-dependence of the local superfluid density can be obtained from EMT \cite{caprara:2013} and assuming proportionality to the measured critical current results in the dashed line in Fig. \ref{fig:ivcurves}(d). The deviation from the data is can be attributed to the connectivity of the array not being accounted for in the model. Fitting to the theory for a junction of arbitrary transparency \cite{KO, Haberkorn:1978} and using the BCS result for the temperature dependence of the gap $\Delta (T)$\cite{richter_interface_2013}, we find good agreement with the data for an individual metallic weak link in the dirty limit, with $J_{JJ} = \frac{\pi \Delta(T)}{2 e R_{JJ}} \tanh{\frac{\Delta(T)}{2k_B T}}$. This fit is shown by the solid line in Fig. \ref{fig:ivcurves}(d) with $T_c = 379 \, \mathrm{mK}$, close to the value $T_c =358 \, \mathrm{mK}$ found from Fig.\ \ref{fig:ivcurves}(a). The difference is within the width of the distribution $\gamma = 24 \, \mathrm{mK}$.

The critical current density as a function of magnetic field is shown in Fig.\ \ref{fig:ivcurves}(e). At high B-field values the sample does not reach the superconducting state, however, a clear transition to a higher resistive state is still observed at a distinct current $I_c$. For an individual uniform rectangular JJ, a magnetic field will cause the measured critical current to oscillate and follow the Fraunhofer pattern $I_c \propto | \mathrm{sin} (\pi \Phi/\Phi_0)/(\pi \Phi/\Phi_0)|$ \cite{tinkham}. For a sample composed of a random array of junctions, the oscillations average out and we expect $J_c(B)$ to follow the approximate envelope $\propto 1/(1+B/B_0)^{\beta}$ where $\beta \approx 1$ depends on the junction geometry \cite{van_der_laan_2001,muller_critical_current_1991} and the characteristic scale $B_0$ relates to the average junction area $A_0 = \Phi_0/\pi B_0$. As seen in Fig.\ 2(e) this simple model shows good overall agreement the data yielding $\beta$ = 1.42 and a junction area of $0.029 \pm 0.002 \, \mathrm{\mu m}^{2}$.

In addition to the measurements presented in Fig.\ \ref{fig:ivcurves} the temperature dependence of the critical current was measured at finite $B$. The resulting superconducting phase diagram of the a-LAO/STO interface is shown in Fig.\ 3. \footnote{To obtain the 3D surface grid, the $J_{c}$ data points obtained from the $IV$-curves were fitted to theory and smoothed using a Hanning window and connected by interpolation.}

\begin{figure}
\includegraphics[width=7cm]{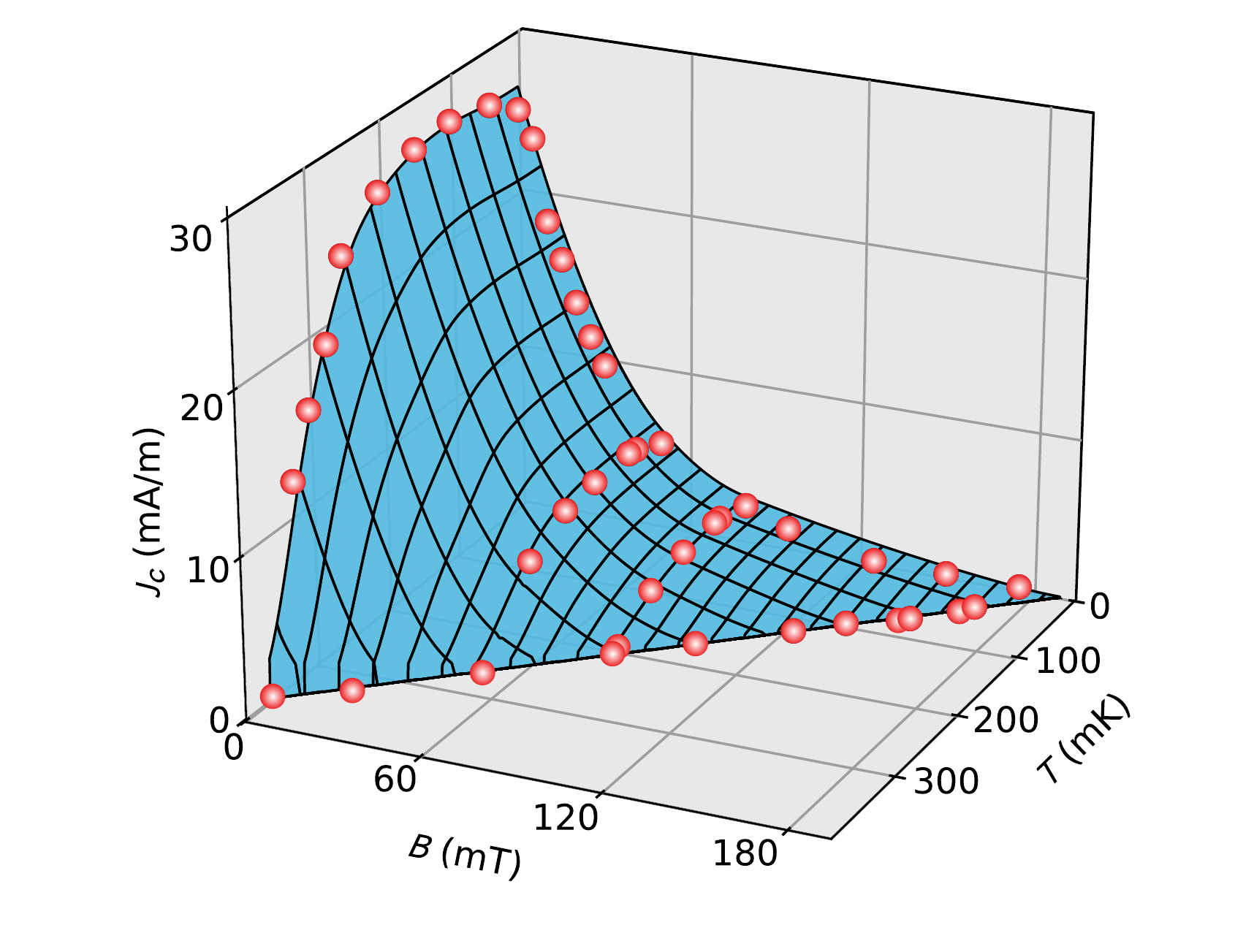}
\caption{\label{fig:phasediagram} Full superconducting phase diagram obtained from the dependence of the critical current density on temperature and magnetic field. The surface grid was obtained from interpolation between data points from Fig. \ref{fig:hallbar}(b), \ref{fig:ivcurves}(d) and (e) (red). Additional $IV$-curve measurements at static magnetic fields have been added as well to establish the shape at the interior of the surface. The superconducting region is shaded in blue.}
\end{figure}

\begin{figure}
\includegraphics[width=8.5cm]{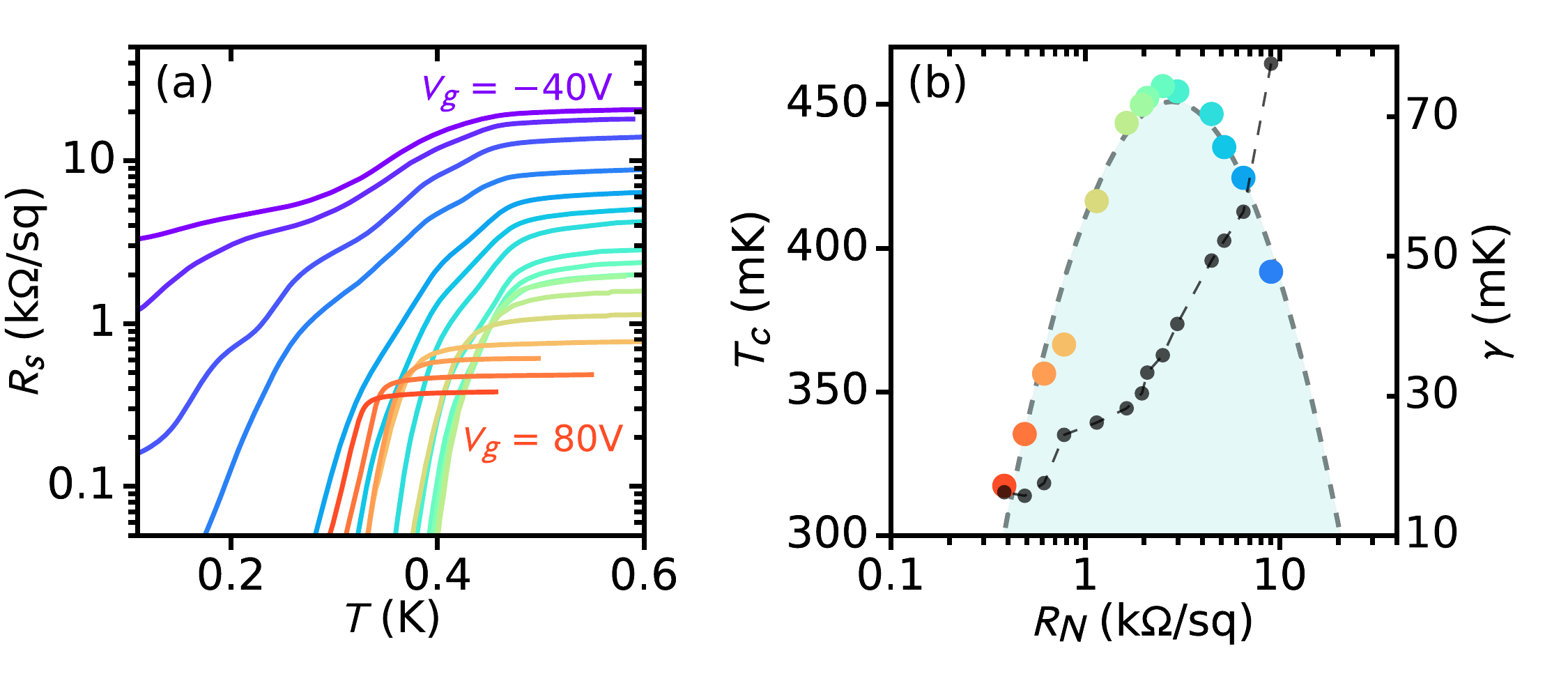}
\caption{\label{fig:dome} (a) Dependence of the sheet resistance $R_s$ on temperature for varying backgate voltages. (b) The critical temperature $T_{c}$ and width of the transition $\gamma$ extracted from the curves in (a) using Effective Medium Theory.}
\end{figure}

A key feature of the superconducting phase in STO-based interfaces is the dome-shaped dependence of the critical temperature on electrostatic doping \cite{caviglia:2008}, related to the doping dependence of $T_c$ in bulk STO \cite{Schooley:1965,Lin:2014}. Figure \ref{fig:dome}(a) shows $R_{s}(T)$ for various $V_{bg}$ for the a-LAO/STO heterostructure and Fig.\ \ref{fig:dome}(b) shows the corresponding $T_c(V_{bg})$-dome extracted using EMT, which reaches a maximum $T_{c}$ of $460 \, \mathrm{mK}$ at optimal doping. This value is larger than what has previously been reported for LAO/STO-based interfaces and very close to the reported transition temperature of bulk conducting STO at optimal doping \cite{Schooley:1965}. At the lowest $V_g$ (highest $R_N$) the resistance does not fall to zero and $T_c$ cannot be defined for these curves. As seen in Fig.\ \ref{fig:dome}(b) the width of the transition $\gamma$ decreases monotonically across the dome. At low $V_{bg}$ (high $R_N$) the $R_{s}(T)$-curves develop multiple steps, which we ascribe to different regions of the sample entering the superconducting state at different temperatures. This is consistent with scanning probe measurements\cite{Bert:2012}, which report a pronounced spatial inhomogeneity in the diamagnetic screening on the underdoped side of the dome.

%

%
\begin{figure}
\includegraphics[width=8.5cm]{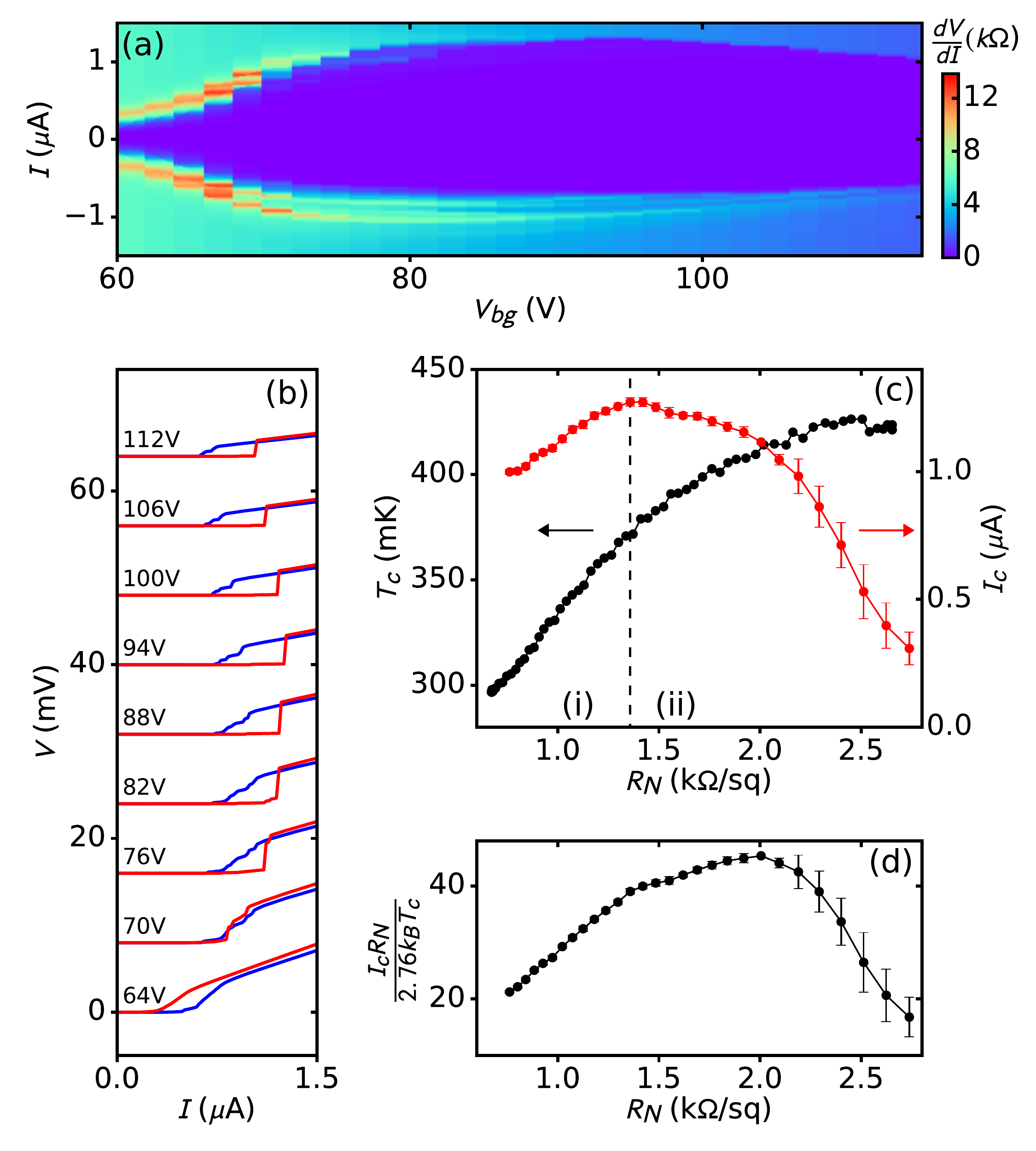}
\caption{\label{fig:compare_Ic_and_Tc} (a) AC differential resistance measured in a separate cooldown for varying DC bias current (from negative to positive values) and backgate voltages. (b) Simultaneously measured DC $IV$-characteristics for varying backgate voltages, plotted with a horizontal offset of $5 \, \mathrm{mV}$ between curves. The bias current is increased in the positive (negative) direction for the red (blue) curve. (c) The critical temperature $T_c$ (black) and the measured critical current $I_c$ (red), extracted from the $IV$-characteristics in (b), versus the normal-state resistance $R_{N}$ (black). (d) The corresponding $I_c R_N$ product. Error bars in (c-d) depict the width of the switching region in the $IV$-characteristics shown in (b).}
\end{figure}

Further insight into the superconducting phase emerges by comparing the backgate dependence of the critical current and critical temperature. Figures \ref{fig:compare_Ic_and_Tc} (a-b) show the $IV$-characteristics at various $V_{bg}$ obtained in a separate cooldown. At each backgate voltage, DC current-biased $IV$-curves and $\partial V / \partial I_{sd}$ were simultaneously measured, from which $I_c(V_{bg})$ is extracted. The gate-dependence of the $T_c$ was obtained using a temperature feedback loop keeping $R_s$ at 50\% of the normal-state resistance.

Figure \ref{fig:compare_Ic_and_Tc}(c) shows $I_c ( V_{bg})$ and $T_c(V_{bg})$ with respect to $R_N$ to compensate for gate hysteresis. Both $T_c$ and $I_c$ exhibit dome-shaped dependencies on doping, however, the two domes peak at significantly different doping levels. Two regimes can be identified: $i$ (for $R_N \lesssim 1.4 \, \mathrm{k\Omega/sq}$) and $ii$ (for $\gtrsim 1.4 \, \mathrm{k\Omega/sq}$). In regime $i$ the device is on the overdoped side of the dome and $T_c$ increases with $R_N$. The critical current $I_c(R_N)$ qualitatively follows $T_c(R_N)$ and both exhibit an increase with $R_N$ with a decreasing rate. At $R_N \approx 1.4 \, \mathrm{k\Omega/sq}$ the critical current peaks at $\sim 1.3 \, \mathrm{\mu A}$ and in regime $ii$, $I_c$ decreases with $R_N$ while $T_c$ continues to increases until it peaks at $R_N \approx 2.5 \, \mathrm{k\Omega/sq}$.

For a conventional homogeneous thin film superconductor, $I_c$ is described with Ginzburg-Landau theory by $I_c^{GL} \propto \Delta \propto T_c$ and is expected to follow $T_c$, unlike the experiment. In the alternative scenario of a Josephson-coupled array, as a simplest model, a single Josephson junction in the superconducting percolation path triggers the transition from the superconducting to the resistive state. The low-temperature critical current is in this case $I_c^{JJ} \propto \Delta/e R_{JJ}$ with $e$ being the electron charge and we assume that $R_{JJ}$ depends on electrostatic doping qualitatively similar to $R_N$. Thus on the overdoped side of the $T_c$-dome, since $V_g$ tunes both $T_c$ and $R_N$, the Josephson-array scenario allows for a situation where an increase in $T_c$ is accompanied by a decrease of $I_c$ as observed in regime $ii$. Here $I_c$ is progressively suppressed as $R_N$ increases and domains are decoupled, while $T_c$ does not depend on the coupling between domains but rather on the carrier density of the individual domains. Therefore this behavior provides a qualitative distinction between the homogeneous thin-film and Josephson-array scenarios, and shows that the latter describes the a-LAO/STO interface superconductivity. The formation of superconducting weak links is also described as the onset of 'weak superconductivity' and is related to the formation of a pseudogap, shown to occur in the normal state of LAO/STO and high-$T_c$ superconductors \cite{richter_interface_2013, Talantsev:2015}. Note that the scaling analysis shown in previous work \cite{caviglia:2008, caprara:2013,biscaras_multiple_2013} to capture $T_c(V_{bg})$ close to the phase transition is also expected to be valid for the transition in a JJA driven by a coupling constant\cite{Sondhi:1997}.

The $IV$-curves shown in Fig. \ref{fig:compare_Ic_and_Tc}(b) exhibit an increasing amount of switching events and a decreasing amount of hysteresis with lowering $V_{bg}$. This behavior is also consistent with a JJA, in the case of a dominating contribution to the $Q$-factors from lowering of the mutual capacitance as the superconducting domains are progressively decoupled and the distribution of critical currents is broadened. For an individual JJ, $I_c R_{JJ} \propto \Delta$ is a constant, and correspondingly for a regular $N \times N$ array, $I_c R_N^{array} \approx N \frac{\pi}{2} \Delta/e \approx N \frac{\pi}{2} 1.76 k_B T_c/e$ is also expected to be constant for a static array. Using the measured $T_c$, equal to the average $\bar{T_c}$ of the distribution according to EMT and assuming $R_N = R_N^{array}$, Fig.\ \ref{fig:compare_Ic_and_Tc}(d) shows the extracted $N$ as a function of doping which follows a dome-like structure peaked at $R_N \approx 1.6 \, \mathrm{k\Omega/sq}$ intermediate between the center values of $I_c (R_N)$ and $T_c (R_N)$. The varying $N$ suggests a gate-dependent structure of the array, possibly related to the doping dependence of the spatial variations observed in scanning probe experiments\cite{Kalisky:2013, Honig:2013}.

In conclusion, we have demonstrated superconductivity in patterned a-LAO/STO and established the superconducting phase diagram. The room-temperature grown top layer enhances the feasibility of conventional micro fabrications techniques for designing gateable mesoscopic superconducting oxide devices. The characteristics of the phase diagram are qualitatively consistent with previous studies of c-LAO/STO samples. We recover the dome-like dependence of the critical temperature on backgate voltage with a peak value of $460 \, \mathrm{mK}$, significantly larger than observed for the c-LAO/STO system and close to the value for bulk conducting STO at optimal doping. From the critical current phase diagram, the observation of multiple resistance steps in the $IV$-characteristics and the observation of a pronounced shift between the $T_c (V_{bg})$ and $I_c(V_{bg})$ domes, we show that the system can be consistently described by a model of an intrinsic Josephson junction array formed by a random network of weakly coupled superconducting domains. The inhomogeneity could be related to inhomogeneous carrier doping by oxygen vacancies or the tetragonal domain boundaries in the STO crystal. The results highlight the important role of inhomogeneity for the properties of superconductivity in STO-based heterostructures. 

\begin{acknowledgments}
We thank A. Smith, M. von Soosten, J. Folk, S.L. Folk, K.C. Nowack, N. Scopigno and H. Suominen for useful discussions, S. Upadhyay, N. Payami, C.B. S{\o}rensen and J. Geyti for technical assistance. R. Egoavil and N. Gauquelin, University of Antwerp are acknowledged for  HAADF STEM imaging. J. Kleibeuker, G. Koster, and G. Rijnders, MESA+ Institute at the University of Twente are acknowledged for input in the discussion. Work was supported by the Lundbeck Foundation and the Danish Agency for Science, Technology and Innovation. The Center for Quantum Devices is supported by the Danish National Research Foundation.
\end{acknowledgments}
\bibliography{refs_nourl}
\end{document}